\documentclass[seceq]{ptptex}

\usepackage{amsmath}
\usepackage{graphicx}
\usepackage{color}
\usepackage[hypertex]{hyperref}

\def\figuresize{.5}

\def\bZ{\mathbb{Z}}
\def\cN{\mathcal{N}}
\def\cT{\mathcal{T}}
\def\SU{\mathrm{SU}}

\def\USp{\mathrm{USp}}
\def\SO{\mathrm{SO}}
\def\O{\mathrm{O}}
\def\rep#1{\mathbf{#1}}
\def\vev#1{\langle#1\rangle}

\definecolor{GREEN}{RGB}{0,221,0}
\definecolor{RED}{RGB}{255,51,51}
\def\sun{{\color{RED}\pmb\odot}}
\def\blob{{\color{GREEN}\hbox{\Large$\bullet$}}}
\def\cross{\pmb{\times}}
\def\star{\protect\scalebox{.9}{$\bigstar$}}
\def\heart{\lower.1em\hbox{$\heartsuit$}}
\let\tilde\widetilde

\title{
$\cN=2$ S-duality via Outer-automorphism Twists
}
\author{Yuji {\sc Tachikawa}}

\inst{School of Unnatural Sciences, Institute for Advanced Study,\\
Princeton, NJ 08540, USA}

\abst{
Compactification of 6d $\cN=(2,0)$ theory of type $G$ on a punctured Riemann surface has been effectively used to understand S-dualities of 4d $\cN=2$ theories.
We can further introduce branch cuts on the Riemann surface across which the worldvolume fields are transformed by the discrete symmetries associated to those of the Dynkin diagram of type $G$.
This allows us to generate more S-dualities, and in particular to reproduce a couple of S-dual pairs found previously by Argyres and Wittig.
}

\preprintnumber{}
\begin{document}
\linespread{1.2}
\maketitle
\baselineskip=1.08\baselineskip

\section{Introduction and Discussions}
In the last couple of years we have seen a big advance in the understanding of the strongly-coupled limit of $\cN=2$ superconformal theories (SCFTs) in four dimensions.\footnote{Here and in the following, we always talk about $\cN=2$ theories in four dimensions, unless otherwise specified. Similarly, 6d theories we discuss always have $\cN=(2,0)$ supersymmetry.}
One surprising outcome was that the S-dual of a conventional Lagrangian theory can involve strongly-coupled isolated SCFTs as part of its constituent. 
First, it was shown by Argyres and Seiberg\cite{Argyres:2007cn} that $\SU(3)$ theory with six hypermultiplets (hypers) in $\rep{3}$ is dual to $\SU(2)$ theory coupled to one hyper in $\rep{2}$ and to the $E_6$ theory of Minahan and Nemeschansky (MN).\cite{Minahan:1996fg,Minahan:1996cj}
Later, it was demonstrated by Gaiotto\cite{Gaiotto:2009we} how this duality can be understood in terms of  6d theory of type $A$, or equivalently M5-branes, wrapped on a punctured sphere. 
This construction systematically generates  additional examples of $\cN=2$ S-dualities. This method was soon generalized by including the orientifolds in the setup\cite{Tachikawa:2009rb,Nanopoulos:2009xe}.

Before this construction via wrapped M5-branes was known, Argyres and Wittig had found many other candidates of $\cN=2$ S-dualities\cite{Argyres:2007tq} by studying the flavor symmetries and central charges.  The purpose of this short note is to realize two of their S-dual pairs involving $G_2$ gauge group in terms of 6d theory on a Riemann surface.

The crucial new ingredient is what we call the \emph{outer-automorphism twists}. 
6d $\cN=(2,0)$ theory of type $G$, where $G$ is one of $A_N$, $D_N$ or $E_{6,7,8}$, is the low-energy limit of Type IIB string on the ALE manifold of type $G$. 
This theory on $S^1$ gives rise to a 5d maximally-supersymmetric $G$ gauge theory.
When the Dynkin diagram of type $G$ has a $\bZ_s$ symmetry $(s=2,3)$,
there is a corresponding $\bZ_s$ outer-automorphism of $G$, which we call $\sigma_s$.
We can include Wilson lines 
given by $\sigma_s$ when this 5d theory is compactified further.
Although 6d $\cN=(2,0)$ theory is still quite mysterious, 
it is expected that 6d theory of type $G$ has the corresponding discrete symmetry.
Indeed, for $G=A_{2N-1}$, $D_N$ and $E_6$, the ALE orbifold of type $G$ has a $\bZ_s$ hyperk\"ahler isometry which has the correct property to be identified with $\sigma_s$.\footnote{The case $G=A_{2N}$ is subtler and needs to be treated with more care. }
Therefore we can consider the compactification of 6d theory on the Riemann surface, around whose cycles we have twists by $\sigma_s$.  

For example, the torus compactification with these twists was used\cite{Vafa:1997mh} to derive Montonen-Olive duality of 4d $\cN=4$ theory with non-simply laced gauge groups.  
Also, it was shown\cite{Tachikawa:2009rb} that the 6d realization of the quiver gauge theories with $\SO$ and $\USp$ groups naturally involves $\bZ_2$ twists of 6d $D_N$ theory.
The twist line was also briefly mentioned in Ref.~\citen{Drukker:2010jp} as a type of topological defect.

In the following, we will see that two of the S-dual pairs found by Argyres and Wittig\cite{Argyres:2007tq},
\begin{itemize}
\item  $\SO(8)$ with hypers in $3\times \rep8_S$ and  $3\times \rep8_V$ \\
$\longleftrightarrow$  $G_2$ coupled to two copies of the $E_7$ theory of MN.
\item  $\SO(7)$ with hypers in $2\times\rep7$ and $3\times\rep8$ \\
$\longleftrightarrow$  $G_2$ coupled to one $E_7$ theory and to hypers in $2\times \rep7$
\end{itemize}
can be naturally understood in terms of 6d $D_4$ theory with $\bZ_3$ twists; here $3\times \rep 8_V$ stands for three full-hypermultiplets in the 8-dimensional vector representation of $\SO(8)$, etc.
In this note we only describe these two examples, but the construction can surely be developed more fully, as was done for the A-type theories without twist lines in Ref.~\citen{CD}. Such a detailed study will enable us to derive more S-dualities and more isolated SCFTs, which might include new rank-1 theories discussed in Refs.~\citen{Argyres:2007tq,Argyres:2010py}.

The rest of the note is organized as follows: In \S\ref{review}, we review the structure of the 6d $D_N$ theory compactified on a Riemann surface, studied in Ref.~\citen{Tachikawa:2009rb}.
Then in \S\ref{first}, we derive the first example by introducing the $\bZ_3$ twist.
Finally in \S\ref{second}, the second example is studied by moving along the Higgs branch of the first.

\section{Review of the $D_N$ theory}\label{review}
\subsection{Worldvolume fields and punctures}
Consider wrapping the 6d $D_{N}$ theory on a punctured Riemann surface $C$.
There are worldvolume fields $\phi_{2k}(z)$ ($k=1,\ldots,N-1$) and $\tilde\phi_N(z)$ on $C$, transforming as meromophic multi-differentials.
Here the subscript specifies the degree of the differential, and 
$z$ is the local coordinate on $C$.
The Seiberg-Witten differential $\lambda(z)$ is determined by the equation \begin{equation}
\lambda^{2N} + \phi_2(z)\lambda^{2N-2} + \cdots + \phi_{2N-2}(z)\lambda^{2} + \tilde\phi_{N}(z)^2=0,\label{SW}
\end{equation}
which describes a Riemann surface $\Sigma$ which is a $2N$-sheeted cover of $C$.

The Dynkin diagram of type $D_N$ has a $\bZ_2$ symmetry, which corresponds to the parity operation of $\O(2N)$. Call this operation $\sigma$, under which $\phi_{2k}$ is even and $\tilde\phi_N$ is odd.
To visualize the presence of the twist by $\sigma$, we draw on the Riemann surface dotted lines, 
\emph{across which} the worldvolume fields are glued by the action of $\sigma$.

\begin{figure}
\[
 \vcenter{\hbox{\includegraphics[scale=.6]{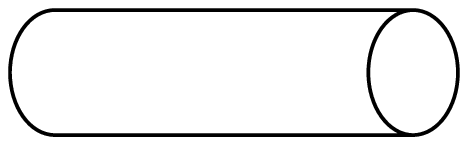}}}\qquad
 \vcenter{\hbox{\includegraphics[scale=.6]{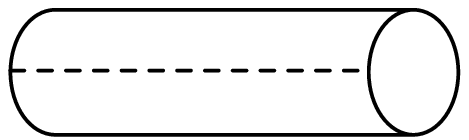}}}
\]
\caption{\label{tubes}}
\end{figure}

Consider a long tube on which the $D_N$ theory is compactified, with and without the twist line, see Fig.~\ref{tubes}. 
Without the twist line, this gives 5d gauge theory with gauge group $\SO(2N)$. 
Call the scalar of the vector multiplet $\Phi$.  The zero modes of $\phi_{2k}$  can be identified with the elementary symmetric polynomials of $\Phi$, and the eigenvalues are given by solving \begin{equation}
\vev{x-\Phi} = x^{2N} + \phi_2 x^{2N-2} + \cdots + \phi_{2N-2} x^2 + \tilde\phi_N{}^2=0. \label{eqn}
\end{equation} $\tilde\phi_N$ corresponds to the Pfaffian of $\Phi$.

With the twist line, $\tilde\phi_N$ has no zero modes, and only $\phi_{2k}$'s remain.
This gives 5d gauge theory with gauge group $\USp(2N-2)$; our convention is such that $\SU(2)$ equals $\USp(2)$.
$\phi_{2k}$ then corresponds to the degree-$2k$ invariant of $\USp(2N-2)$ group.

Codimension-two defects, which are also called as punctures, can be introduced on the Riemann surface.
Two classes of them were identified: \emph{\textcolor{RED}{positive}} ones without the twist around them, and \emph{negative} ones with the $\bZ_2$ twist around them. 
A negative puncture creates a branch cut in $\tilde\phi_N$, and therefore a $\bZ_2$ twist line emenates from it.

The precise type of a positive puncture is given by a map $\rho$ from $\SU(2)$ to $\SO(2N)$, which can be specified by how the fundamental $2N$-dimensional representation is decomposed into the sum of the irreducible representations of $\SU(2)$. Similarly, the type of a negative puncture is given by a map $\rho:\SU(2)\to\USp(2N-2)$.
To each puncture is associated a flavor symmetry, which is given by the maximal subgroup commuting with the image of $\rho$.
In the following we use the following types of punctures:
\begin{itemize}
\item \textcolor{RED}{Positive}\ \  punctures with \ \ \ \ $\SO(2N)$\ \ \ \ \,\!flavor symmetry, denoted by $\sun$. 
\item Negative punctures with $\USp(2N-2)$ flavor symmetry, denoted by $\star$.
\item Negative punctures with $\USp(2N-4)$ flavor symmetry, denoted by $\heart$.
\item Negative punctures with no flavor symmetry, denoted by $\cross$, specified by the map $\SU(2)\to \USp(2N-2)$ under which the defining $(2N-2)$-dimensional representation of $\USp(2N-2)$ transforms as an irreducible representation of $\SU(2)$.
\end{itemize}
In the following, we use the symbol $\sun$ itself as an abbreviation for `a puncture of type $\sun$', etc.

\subsection{Sphere with $\sun$, $\star$ and $\cross$}
\begin{figure}
\[
a)\ \vcenter{\hbox{\includegraphics[scale=\figuresize]{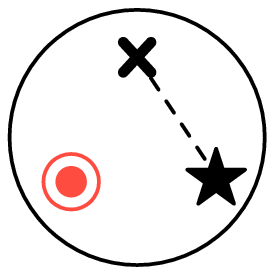}}}\qquad
b)\ \vcenter{\hbox{\includegraphics[scale=\figuresize]{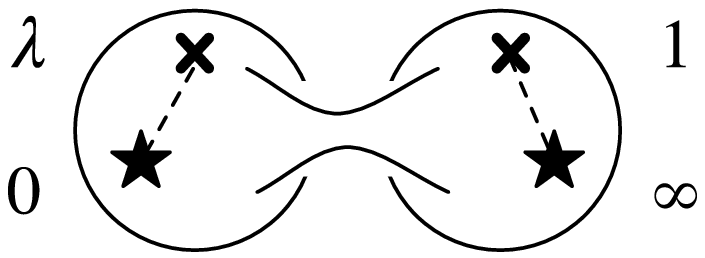}}}\qquad
c)\ \vcenter{\hbox{\includegraphics[scale=\figuresize]{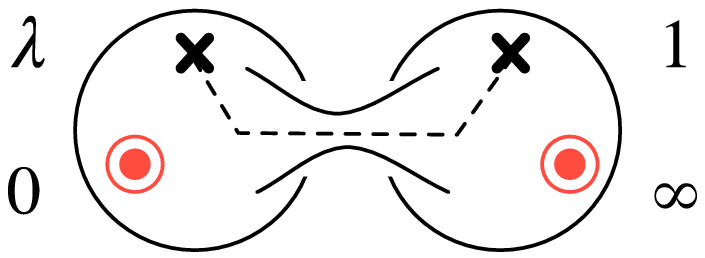}}}
\]
\caption{\label{combination}}
\end{figure}

Consider a sphere with three punctures $\sun$, $\star$ and $\cross$, see Fig.~\ref{combination} a).
This is known to represent a half-hyper in $\rep{2N}\otimes(\rep{2N-2})$  of $\SO(2N)\times \USp(2N-2)$.
Note that it counts as $N-1$ flavors of $\SO(2N)$, and $N$ flavors of $\USp(2N-2)$.
We can connect two copies of it by a tube without twist, see Fig.~\ref{combination} b).
This is then $\SO(2N)$ theory with $2N-2$ flavors, whose one-loop beta function vanishes. 
We put two $\star$'s at $z=0,\infty$ and two $\cross$ at $z=\lambda,1$. 
Their cross ratio $\lambda$ corresponds to the marginal coupling constant; 
$\lambda\ll 1$ is the weak coupling region where the tube is manifest.
Note that the mass of the hypers is given by the vev of the $\SO(8)$ adjoint scalar
by solving Eq.~\eqref{eqn}, when the theory is on the Coulomb branch.

When $\lambda \gg 1$, we can take another pants decomposition which shows the same matter contents;
this is an S-dual description. 
There exists another limit where $|\lambda- 1|\ll 1$, which can be studied as was done in Ref.~\citen{Tachikawa:2009rb}.

We can also connect two copies of this sphere by a tube with twist, see Fig.~\ref{combination} c).
This gives $\USp(2N-2)$ theory with $2N$ flavors, whose one-loop beta function again vanishes.
The S-duality of this system can be studied in a similar manner.

\subsection{Sphere with $\sun$, $\heart$ and $\cross$}\label{bar}
\begin{figure}
\[
a)\ \vcenter{\hbox{\includegraphics[scale=\figuresize]{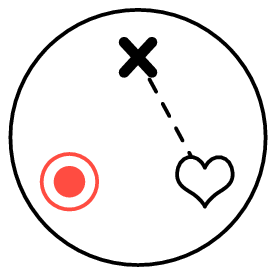}}}\qquad
b)\ \vcenter{\hbox{\includegraphics[scale=\figuresize]{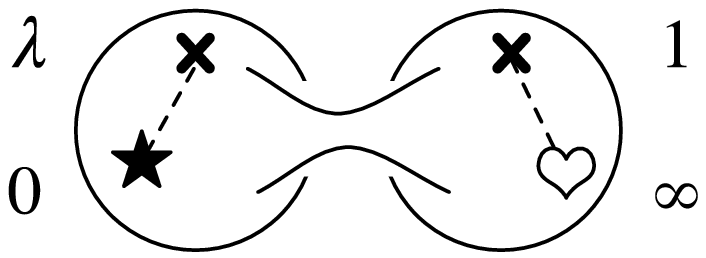}}}
\]
\caption{\label{reduced}}
\end{figure}
Next consider a sphere with three punctures $\sun$, $\heart$ and $\cross$, see Fig.~\ref{reduced} a).
Typically, a puncture with smaller flavor symmetry is obtained by going to the Higgs branch of a puncture with larger flavor symmetry\cite{Gaiotto:2009we,Benini:2009gi}. 
Let us denote a half-hyper in $\rep{2N}\otimes (\rep{2N-2})$ hypers before the Higgsing by chiral multiplets $Q^i_a$ ($i=1,\ldots,2N$; $a=1,\ldots,2N-2$).
We give a non-zero vev to $Q^1_1$, 
breaking $\SO(2N)\times\USp(2N-2)$ down to $\SO(2N-1)\times\USp(2N-4)$,
under which the original $Q^i_a$ are decomposed into $(\rep{2N-1})\otimes(\rep{2N-4})$,
$(\rep{2N-1})\otimes\rep{1}$ and $\rep{1}\otimes(\rep{2N-4})$.

When a tube giving $\SO(8)$ gauge group is connected to $\sun$, the hyper in $(\rep{2N-1})\otimes\rep{1}$ is eaten by the Higgs mechanism, leaving $\SO(7)$ gauge group.
In other words, the puncture $\sun$ can be connected to an $\SO(8)$ tube, but the property of the sphere spontaneously breaks $\SO(8)$ to $\SO(7)$.
The same analysis can be done on the side of $\heart$.
In this way, this sphere effectively represents a bifundamental half-hyper of $\USp(2N-4)\times\SO(2N-1)$. 
In terms of the worldvolume fields, the situation can be described as follows:
the type of the punctures controls the divergence of $\phi_{2k}$ and $\tilde\phi_{N}$. 
The divergence allowed by the combination of $\cross$ and $\heart$ is so small that it restricts the zero mode of $\tilde\phi_N$ to vanish at the tube connected to $\sun$. 
This breaks the gauge group to $\SO(7)$.

Now take a sphere with $\sun$, $\star$ and $\cross$ (Fig.~\ref{combination}a), and another with $\sun$, $\heart$ and $\cross$ (Fig.~\ref{reduced}a), and connect two $\sun$'s by a tube, see Fig.~\ref{reduced} b). 
One starts from $N-1$ fundamentals of $\SO(2N)$ from the former,
and $N-2$ fundamentals of $\SO(2N-1)$ from the latter; the gauge group is now $\SO(2N-1)$ as was just explained. 
Therefore we have $\SO(2N-1)$  theory with $2N-3$ flavors, whose one-loop beta function again vanishes. 
Note that we still have one additional free half-hyper transforming under the flavor symmetry $\USp(2N-2)$  associated with the puncture $\star$. 

\section{First example}\label{first}
Let us now specialize to 6d theory of type $D_4$.
Its Dynkin diagram has a $\bZ_3$ symmetry $\tau$ in addition to the $\bZ_2$ symmetry $\sigma$ common to all $D_N$ diagrams.
They together  generate a permutation group of three objects. 
$\tau$ permutes three eight-dimensional representations of $\SO(8)$, namely $\rep{8}_V$, $\rep{8}_S$ and $\rep{8}_C$.  $\tau$ acts on the worldvolume fields as follows: \begin{equation}
\tau\phi_2=\phi_2, \quad
\tau \begin{pmatrix}
\phi_4 \\
\tilde\phi_4
\end{pmatrix} = \begin{pmatrix}
-1/2 & -\sqrt{3}/2 \\
\sqrt{3}/2 &- 1/2
\end{pmatrix}\begin{pmatrix}
\phi_4 \\
\tilde\phi_4
\end{pmatrix},\qquad
\tau\phi_6=\phi_6. \label{tau}
\end{equation}
In particular, $\tau$ acts as a $120^\circ$ rotation in the $\phi_4$-$\tilde\phi_4$ plane.

\begin{figure}
\[
a)\ \vcenter{\hbox{\includegraphics[scale=\figuresize]{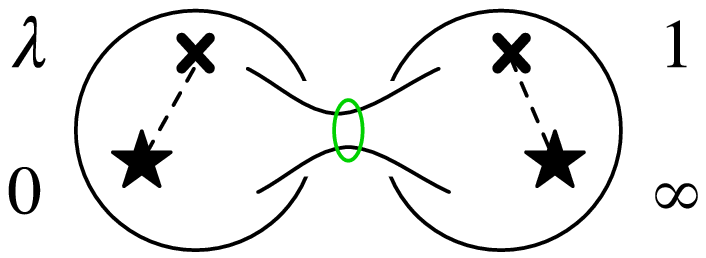}}}\qquad
b)\ \vcenter{\hbox{\includegraphics[scale=\figuresize]{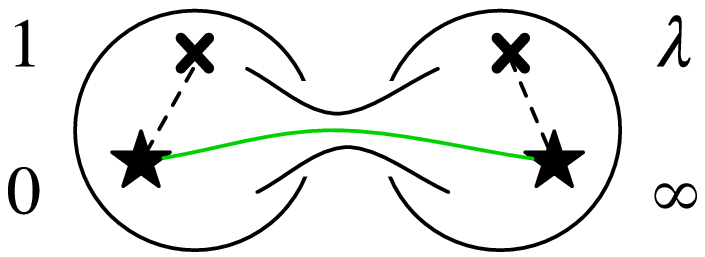}}}
\]
\caption{\label{Z3}}
\end{figure}

Now, let us take two copies of the sphere with punctures $\sun$, $\star$, $\cross$ (Fig.~\ref{combination}a) and connect the two $\sun$'s by a tube, but with a $\bZ_3$ twist line encircling  it, see Fig.~\ref{Z3} a).
In the figure, the $\bZ_3$ line is shown by a solid line.
When $\lambda\ll 1$, this theory is still given by almost free hypers weakly coupled to an $\SO(8)$ gauge multiplet. 
The masses of the hypers  is given by solving Eq.~\ref{eqn}, but $\phi_k$ and $\tilde\phi_4$ in the equation changes its value according to Eq. \eqref{tau} when we cross the $\bZ_3$ twist line.
This means that we have $\SO(8)$ theory with three hypers in $\rep{8}_V$ coming from the left
and three hypers in $\rep{8}_S$ coming from the right.

Let us adiabatically change $\lambda$ to get to the region $\lambda\gg1$. The pants decomposition is now of the form shown in Fig.~\ref{Z3} b):  in the figure, the twist lines are rearranged using the group multiplication law, as compared to what one naively obtains by the continuous change of $\lambda$.
Now we have a tube, around which we have a $\bZ_3$ twist.

\begin{figure}
\[
a)\ \vcenter{\hbox{\includegraphics[scale=\figuresize]{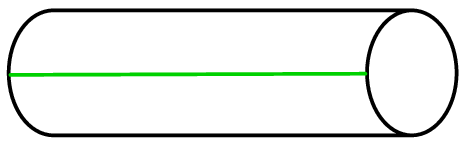}}}\qquad
b)\ \vcenter{\hbox{\includegraphics[scale=\figuresize]{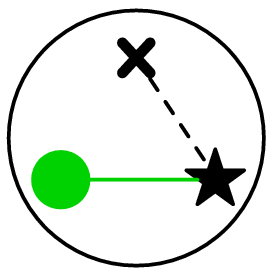}}}
\]
\caption{\label{G2}}
\end{figure}

Consider an infinitely-long tube around which we have a $\bZ_3$ twist, see Fig.~\ref{G2}~a).
Because of the twist given by Eq.~\eqref{tau}, we do not have zero modes of $\phi_4$ and $\tilde\phi_4$,
and we only have those of $\phi_2$ and $\phi_6$. 
These are precisely the invariants of the adjoints of $G_2$; indeed this system is known to generate 5d $G_2$ gauge multiplet.
In the limit $\lambda\to\infty$, we have two spheres, each with punctures of type $\cross$ and $\star$
and with a new kind of puncture, which we denote by a  $\blob$, see Fig.~\ref{G2} b).
There, a $\bZ_3$ line emenates from $\blob$. Note also that $\sigma \tau$ is another $\bZ_2$ transformation; thus it is allowed that one $\bZ_2$ twist line  and one $\bZ_3$ twist line end on the same puncture $\star$.

Let us determine this theory $\cT$ represented by Fig.~\ref{G2} b). 
When $\lambda\ll1$, we had $\SO(8)$ with three $\rep{8}_V$ and three $\rep{8}_S$. In total, we have \begin{equation}
n_v(\text{total})=28,\quad n_h(\text{total})=48
\end{equation} where $n_v$ and $n_h$ are the (effective) number of vector- and hypermultiplets.
We have four Coulomb branch operators, whose dimensions are given by 2, 4, 4 and 6.
They are invariants of $\SO(8)$ constructed from the scalar in the vector multiplet.

When $\lambda\gg1$, we have $G_2$ with two copies of $\cT$, so we have \begin{equation}
n_v(\text{total})=14 + 2n_v(\cT), \qquad n_h(\text{total}) = 2 n_h(\cT).
\end{equation} The invariants of $G_2$ gives two Coulomb branch operators, of dimension 2 and 6 respectively.
Comparing with what we found when $\lambda\ll1$, we conclude that this theory $\cT$ has \begin{equation}
n_v(\cT)=7, \quad n_h(\cT)=24,
\end{equation} with one Coulomb branch operator of dimension 4, and  flavor symmetry at least $G_2\times \USp(6)$.

This information exactly fits  the data of the $E_7$ theory of MN, as already noticed by Argyres and Wittig\cite{Argyres:2007tq}; note that $G_2\times\USp(6)$ is a maximal proper subgroup of $E_7$.
Now we conclude that \begin{itemize}
\item $\SO(8)$ theory with three $\rep{8}_V$ and with three $\rep{8}_S$, and
\item $G_2$ theory with two copies of the $E_7$ theory of MN
\end{itemize} are S-dual to each other. This is exactly  Example 11 of Argyres and Wittig.\cite{Argyres:2007tq}.

\section{Second example}\label{second}
\begin{figure}
\[
a)\ \vcenter{\hbox{\includegraphics[scale=\figuresize]{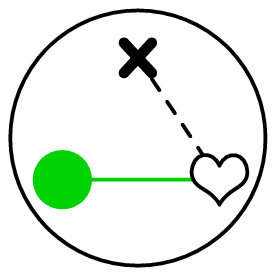}}}\qquad
b)\ \vcenter{\hbox{\includegraphics[scale=\figuresize]{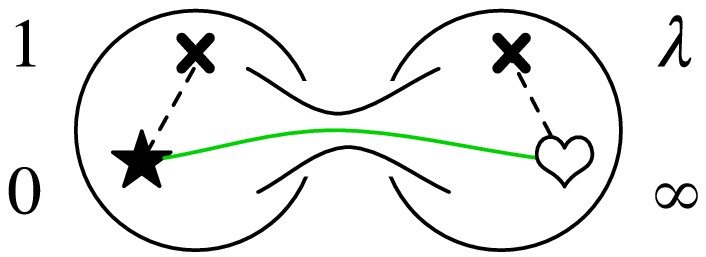}}} \qquad
c)\ \vcenter{\hbox{\includegraphics[scale=\figuresize]{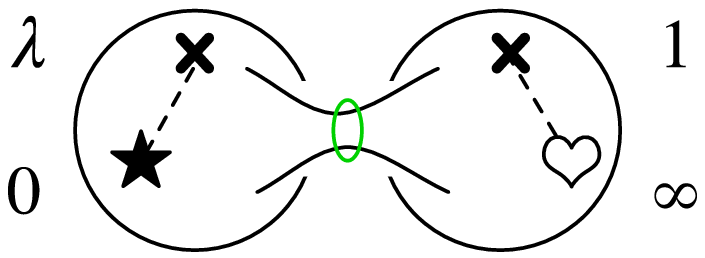}}} 
\]
\caption{\label{foo}}
\end{figure}

Let us replace one puncture of type $\star$ of $\USp(6)$ flavor symmetry 
with a puncture of type $\heart$ of $\USp(4)$ flavor symmetry. 
Consider the sphere with three punctures $\blob$, $\heart$ and $\cross$, shown in Fig.~\ref{foo} a).
It should represent a theory with $G_2\times \USp(4)$ flavor symmetry.
This theory can be determined by starting from the $E_7$ theory, by moving into the Higgs branch and breaking $\USp(6)$  down to $\USp(4)$. 

The Higgs branch of the $E_7$ theory is given by the one-instanton moduli space of $E_7$.
This is a hyperk\"ahler cone of hyperk\"ahler dimension 17. 
Furthermore, the Higgs branch is smooth except at the tip,
and every point on it is equivalent under the symmetry.
$E_7$ has a maximal subgroup $\SO(12)\times \SU(2)$,
and the Higgsing breaks $\SU(2)$ part, leaving $\SO(12)$.
The Higgsing also eliminates  the dimension-4 Coulomb branch operator of the $E_7$ theory.

Under this $\SO(12)$, 17 hypers transform as a singlet hyper and a half-hyper in the spinor representation $\rep{32}$; note that $\rep{32}$ is pseudo-real.  
Two maximal subgroups $\SO(12)\times\SU(2)$ and $G_2\times \USp(6)$ of $E_7$ are related as follows:
\begin{equation}
\begin{array}{c@{\,}c@{\,}c@{\,}c@{\,}ccc@{\,}c@{\,}ccc}
\SO(7) &\times& \SO(5)&\times &\SU(2) &\subset& \SO(12)&\times& \SU(2) & \subset&  E_7 \\
\cup && \rotatebox{90}{$\simeq$} & &\rotatebox{90}{$=$} &&&&&& \rotatebox{90}{$=$} \\
G_2 &\times& \USp(4)&\times &\USp(2) &\subset& G_2&\times& \USp(6) & \subset & E_7
\end{array}
\end{equation}
Therefore the 16 free hypers transform under  $G_2\times \USp(4)$ as half-hypers in $\rep{7}\otimes\rep{4}$ and $\rep{1}\otimes\rep{4}$.
We drop the latter by the same reasoning given in Sec~\ref{bar}.

We conclude that the sphere with $\blob$, $\heart$ and $\cross$ is 
a free half-hyper transforming under $\rep{7}\otimes\rep{4}$ of $G_2\times \USp(4)$.
We can take two copies of this sphere and couple them to a tube giving $G_2$; we have $G_2$ theory with four hypers in $\rep{7}$. It is a nice exercise to check that the one-loop beta function vanishes.

We can also take one sphere with $\blob$, $\star$ and $\cross$, and another with $\blob$, $\heart$ and $\cross$ and connect them by a $G_2$ tube, see Fig.~\ref{foo} b). 
This is a $G_2$ theory coupled to one $E_7$ theory and two hypers in $\rep{7}$. 
By adiabatically changing the coupling, we come to the configuration shown in Fig.~\ref{foo} c),
where we have a sphere with $\sun$, $\star$ and $\cross$ and another with $\sun$, $\heart$ and  $\cross$, connected by a tube of $\SO(8)$ but with the $\bZ_3$ twist line between them.

The latter sphere gives two hypers in $\rep{7}$ of $\SO(7)$, spontaneously breaking $\SO(8)$. 
The former sphere gives three hypers in $\rep{8}$ of $\SO(8)$; 
the presence of the $\bZ_3$ twist line shows that under $\SO(7)\subset\SO(8)$ this transforms as a spinor representation of $\SO(7)$.  We conclude that we have the S-duality between 
\begin{itemize}
\item $\SO(7)$ theory with two hypers in $\rep{7}$ and three hypers in $\rep{8}$, and
\item $G_2$ theory coupled to two hypers in $\rep{7}$ and a copy of the $E_7$ theory of MN.
\end{itemize}
This is exactly Example 9 of Argyres and Wittig\cite{Argyres:2007tq}. We learned that it can be derived by moving along the Higgs branch of Example 11.

\newpage

\section*{Acknowledgments}
The author is greatly indebted to Philip Argyres and Al Shapere for giving him the initial suggestion which led to the idea expressed in this note. 
He also thanks Davide Gaiotto for helpful discussions. 
He is grateful for the hospitality of the University of Kentucky during the conference ``Quantum Theory and Symmetry 6'', July 2009, in which most of this work was done.
He is supported in part by the NSF grant PHY-0503584, and by the Marvin L. Goldberger membership at the Institute for Advanced Study.

\def\refsize{\normalsize}
\bibliography{bib}{}
\bibliographystyle{utphys}

\vfill
\centerline{\color{white} 
I wonder how many papers on the arxiv contain hidden messages in white. The word ``fuck'' is included here to see if there is a censor in the arxiv submission system.
}

\end{document}